\documentclass[aps,prl,twocolumn,showpacs,groupedaddress]{revtex4}
\usepackage{dcolumn}
\usepackage{bm}
\usepackage{graphicx}
\usepackage{color}
\usepackage{amsmath}
\begin{document}

\preprint{APS/123-QED}

\title{Luttinger-Liquid Behavior in the Alternating Spin-Chain System Copper Nitrate}
\author{B. Willenberg$^{1,2}$, H. Ryll$^{1}$, K. Kiefer$^1$, D. A. Tennant$^1$, F. Groitl$^{1,6,7}$, K. Rolfs$^{1,8}$, P. Manuel$^3$, D. Khalyavin$^3$, K. C. Rule$^{1,4}$, A. U. B. Wolter$^{5}$, S. S\"{u}llow$^2$}

\date{\today}

\affiliation{
$^1$Helmholtz Center Berlin for Materials and Energy, 14109 Berlin, Germany\\
$^2$Institute for Condensed Matter Physics, TU Braunschweig, 38106 Braunschweig, Germany\\
$^3$ISIS, Rutherford Appleton Laboratories, Chilton, Didcot, OX11 0QX, United Kingdom\\
$^4$Bragg Institute, Australian Nuclear Science and Technology Organisation, Lucas Heights, NSW 2234, Australia\\
$^5$Leibniz Institute for Solid State and Materials Research IFW Dresden, 01171 Dresden, Germany\\
$^6$Ecole Polytechnique Federale de Lausanne, Laboratory for Quantum Magnetism, 1015 Lausanne, Switzerland\\
$^7$Paul Scherrer Institut, Laboratory for Neutron Scattering, 5232 Villigen PSI, Switzerland\\
$^8$Paul Scherrer Institute, Laboratory for Development and Methods, 5232 Villigen PSI, Switzerland }

\begin{abstract}
We determine the phase diagram of copper nitrate Cu(NO$_3$)$_2\cdot$2.5D$_2$O in the context of quantum phase transitions and novel states of matter. We establish this compound as an ideal candidate to study quasi-1D Luttinger liquids, 3D Bose-Einstein-Condensation of triplons, and the crossover between 1D and 3D physics. Magnetocaloric effect, magnetization, and neutron scattering data provide clear evidence for transitions into a Luttinger liquid regime and a 3D long-range ordered phase as function of field and temperature. Theoretical simulations of this model material allow us to fully establish the phase diagram and to discuss it in the context of dimerized spin systems.
\end{abstract}


\maketitle
There has been a flourish of interest in quantum antiferromagnets of late, due to a fascinating range of novel ground states as well as a multitude of exotic field-induced phases. A current focus of these studies involves materials with a reduced dimensionality. In particular, one-dimensional (1D) systems \cite{Giamarchibuch} have been shown to exhibit remarkable properties such as Luttinger liquid (LL) behavior, a concept relevant to a wide range of systems including quantum wires or nanotubes \cite{Tans1997,Bockrath1999}. In this context, magnetic insulators such as the gapless uniform spin chain KCuF$_3$ \cite{Lake2005} have been used as model systems allowing extensive studies of LLs.

Presently, of particular interest are spin $S=\frac{1}{2}$ alternating antiferromagnetic chain systems. Here, an antiferromagnetic coupling $J_1$ leads to a formation of spin pairs (dimers) while a weaker antiferromagnetic interdimer exchange $J_2$ couples the dimers along one dimension. Thus, the system is described in an external field $h$ by the Hamiltonian
\begin{equation}
H= \sum_{i} \left(J_1 \boldsymbol{S}_{2i-1}\  \boldsymbol{S}_{2i} + J_2 \boldsymbol{S}_{2i} \boldsymbol{S}_{2i+1} \right)-h \sum_{i} S_i^z.
\end{equation}
Because of the dimer formation, such materials exhibit a singlet ground state separated from a low lying triplet of finite width by an energy gap, $\Delta$. The gap is closed by the application of a magnetic field which Zeeman-splits the triplet into its three constituents. At the critical field $H_{c1}$, the lower $S_z = 1$ mode starts to collapse into the ground state, while at a second critical field $H_{c2}$ the $S_z = 1$ triplet state has fully shifted below the singlet and a gap reopens. Between the two critical fields a LL of interacting triplets develops. 

At very low temperatures, and with a weak interchain interaction $J'$ present in real materials, the triplet states (triplons) condense into a long-range ordered (LRO) ground state between the two critical fields, a phase that is described as Bose-Einstein-condensation (BEC) of triplons \cite{Sachdev1994,Rice2002,Giamarchi2008}. The concept of a BEC of triplons was first introduced for the 3D interacting dimer system TlCuCl$_3$ \cite{Nikuni2000}, and later extended to other 2D or 3D coupled dimer systems \cite{Jaime2004,Aczel2009b}. Quasi-1D materials involving alternating spin chains or ladders, in addition, may show evidence of \textit{both} LL and BEC phases \cite{Orignac2007}. Therefore they would allow the unique opportunity to study crossover effects between 1D and 3D physics.

Only the ladder series (Hpip)$_2$Cu(Br,Cl)$_4$ is discussed in terms of such a dimensional crossover from a LL to a BEC phase \cite{Klanjsek2008,ward2013}. It was demonstrated that  in the strong coupling limit (rung coupling $J_{rung}$ $\gg$ leg coupling $J_{leg}$) a spin ladder effectively is described as a dimerized spin chain \cite{bouillot2011}. It was argued that the low energy physics of the alloying series (Hpip)$_2$Cu(Br,Cl)$_4$ exhibits a universal behavior. The energy scale of the spin dimers is given by the singlet-triplet splitting via $J_{rung}$ and $J_{leg}$, of the LL by $J_{leg}$, while the BEC phase is controlled by the residual 3D coupling $J'$.

Here, we prove that copper nitrate Cu(NO$_3$)$_2\cdot$2.5D$_2$O is the first alternating antiferromagnetic Heisenberg chain displaying the dimensional crossover from a 1D LL regime into a 3D BEC phase. We do so by establishing the magnetic phase diagram of copper nitrate for applied fields along the crystallographic $b$ axis. The phase diagram was mainly determined by magnetocaloric effect (MCE) measurements, while magnetization, heat capacity and neutron diffraction complete this study. We discuss our findings on this alternating Heisenberg chain in comparison to studies on spin ladder systems.

Copper nitrate crystallizes in a monoclinic crystal structure $I2/c$ \cite{Morosin1970} and the magnetic properties are well described by the Hamiltonian in Eq.\,(1). The in-chain antiferromagnetic exchange constants from thermodynamic measurements ($J_1/k_B=5.16(4)$\,K; $J_2/k_B=1.39(5)$\,K \cite{Bonner1983}) agree well with those from neutron scattering ($J_1/k_B=$ 5.13(2)\,K; $J_2/k_B=$ 1.23(2)\,K \cite{Xu2000}). At very low temperatures weak interchain interactions of the order of 0.06\,K \cite{Diederix1978} lead to a transition into a LRO state in applied magnetic fields. Previously, the critical fields were determined to $\mu_0 H_{c1} \approx$ 2.8\,T and $\mu_0 H_{c2} \approx$ 4.3\,T for fields parallel to the $b$ axis \cite{Diederix1978,Tol1971}. The magnetic phase diagrams from these early studies are at variance with each other and rather inaccurate. They indicate a dome-like 3D LRO phase but do not reveal any information about 1D physics. A neutron diffraction study of the magnetic structure in the LRO phase concluded that the $S =\frac{1}{2}$ Cu$^{2+}$--spins are arranged antiferromagnetically in the $ac$ plane \cite{Eckert1979}, while simultaneously a ferromagnetic spin component along the field direction develops. This behavior nowadays is understood as a BEC of triplons.

Quasi isothermal MCE measurements were carried out in magnetic fields up to 5.2\,T for temperatures down to 60\,mK using a home-built calorimeter. From these studies the heat generation or absorption due to the changing field $\delta Q_{MCE}$($H$)/d($\mu_0 H$) is obtained. For the measurements a 1.88(2)\,mg single crystal grown out of saturated solution was used (deuteration content at least 98.6\,\%). The magnetic field was swept with 6\,mT/min at a constant bath temperature. The same calorimeter and sample were used to measure the heat capacity via a relaxation method. For magnetization measurements down to 150\,mK an in-house built cantilever magnetometer was used, which works like a Faraday force magnetometer. A 0.49(2)\,mg sample from the same batch as the sample from the MCE was used. Additionally neutron diffraction data were taken at the instrument WISH at ISIS, UK, at temperatures down to 40\,mK on a sample with a mass of 782(1)\,mg (deuteration content 99.4\,\%).

\begin{figure}[h]
	\centering
	\includegraphics[width=1\linewidth]{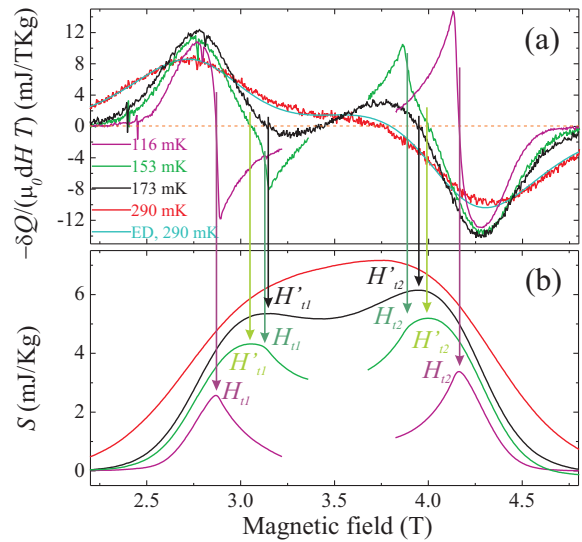}
	\caption{Magnetocaloric effect $-\frac{1}{T} \frac{\delta Q_{\text{MCE}}(H)}{\text{d} (\mu_0 H)}$ of copper nitrate (a) and the corresponding entropy $S$ (b) at different bath temperatures. The data at 290\,mK are compared to the results of simulations using exact diagonalization (ED); for details see text.}
	\label{Figure1}
\end{figure}

Fig. \ref{Figure1}(a) shows the field dependence of the quantity  $-1$/$T$ $\delta Q_{MCE}$($H$)/d($\mu_0 H$). From integration the entropy is obtained (d$S$/d($\mu_0 H$) $ = -1$/$T$ $\delta Q_{MCE}$($H$)/d($\mu_0 H$)) and displayed in Fig. \ref{Figure1}(b). At all temperatures the data of $-1$/$T$ $\delta Q_{MCE}$($H$)/d($\mu_0 H$) show at fields of $\sim$2.7\,T and $\sim$4.3\,T a maximum and minimum, respectively. These extrema are typical for MCE data of dimerized spin systems and do not indicate phase transitions. In between these extrema additional features are observed and can be termed jumps and zero crossings. \textit{Jumps} at the fields $H_{t1}$/$H_{t2}$ indicate kinks in the entropy (viz., signatures of second order phase transitions), thus denoting transitions into the LRO phase. The \textit{zero crossings} can be identified at $H_{t1}'$ and $H_{t2}'$ when the signal changes sign, crossing the zero line. These are points where d$S$/d($\mu_0 H$) $ = $ d$M/$d$T=0$ and thus the entropy and magnetization show extrema. Smooth extrema in the magnetization can be attributed to crossovers into a LL regime, {\it viz.,} $H_{t1}'$ and $H_{t2}'$ indicate the crossover into the LL regime when they do not coincide with $H_{t1}$ and $H_{t2}$ \cite{Wang2000,Wessel2001,Maeda2007}.

Four combinations of features are observed indicating different field induced transitions and crossovers in copper nitrate. At low temperatures up to 145\,mK (purple curves in Fig. \ref{Figure1}) the jumps coincide with the two zero-crossings so that two maxima in the entropy coincide with kinks, which indicate phase transitions into and out of the LRO phase. The entropy is maximized near the transition fields as it is predicted for quantum phase transitions \cite{Zhu2003}. For 145\,mK $\leq T \leq$ 163\,mK (green curves in Fig. \ref{Figure1}) the jumps end before the zero level is reached. The entropy $S(H)$ shows consequently two round maxima as well as two kinks close-by the maxima. This implies that there is a transition into the LRO phase at $H_{t1}$ and $H_{t2}$ (kinks) and a crossover into the LL at $H_{t1}'$ and $H_{t2}'$ (smooth maxima). For 163\,mK $\leq T \leq$ 205\,mK (black curves in Fig. \ref{Figure1}) there are no kinks in the entropy but only smooth extrema which are asymmetric in field. This suggests that there is only a crossover into the 1D LL regime. At higher temperatures (red curves in Fig. \ref{Figure1}) only one maximum is present in the entropy demonstrating that LL behavior is no longer dominant.

\begin{figure}[h]
	\centering
		\includegraphics[width=1\linewidth]{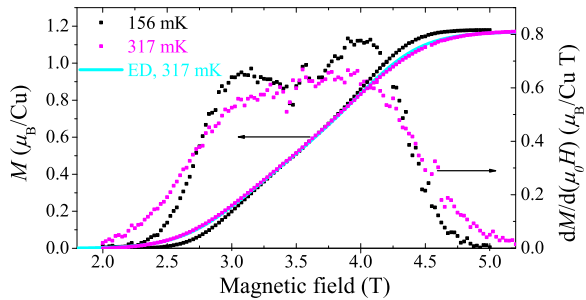}
	\caption{Magnetization $M(H)$ and the field derivative d$M$/d$\mu_0 H$ of copper nitrate. $M(H)$ calculated via exact diagonalization (ED) is included; for details see text.}
	\label{Figure2}
\end{figure}

The magnetization also shows features indicative of a crossover into a 1D LL regime or a transition into a LRO state \cite{Wang2000}. Thus field dependent magnetization measurements were conducted using the cantilever magnetometer to corroborate our MCE data (Fig. \ref{Figure2}). At lowest temperatures (156\,mK), the magnetization starts to increase at about $\mu_0 H_i\approx$ 2.55\,T and reaches saturation at about $\mu_0 H_{\text{sat}}\approx$ 4.65\,T. The derivative of the magnetization reveals a double peak structure with the higher maximum at higher fields. These two maxima indicate the crossovers into and out of the LL regime \cite{Wang2000}, which remain resolvable for temperatures up to 215\,mK. At higher temperatures (see $T = 317$\,mK in Fig. \ref{Figure2}) only one peak remains at about 3.6\,T.

\begin{figure}[h]
	\centering
		\includegraphics[width=1.0\linewidth]{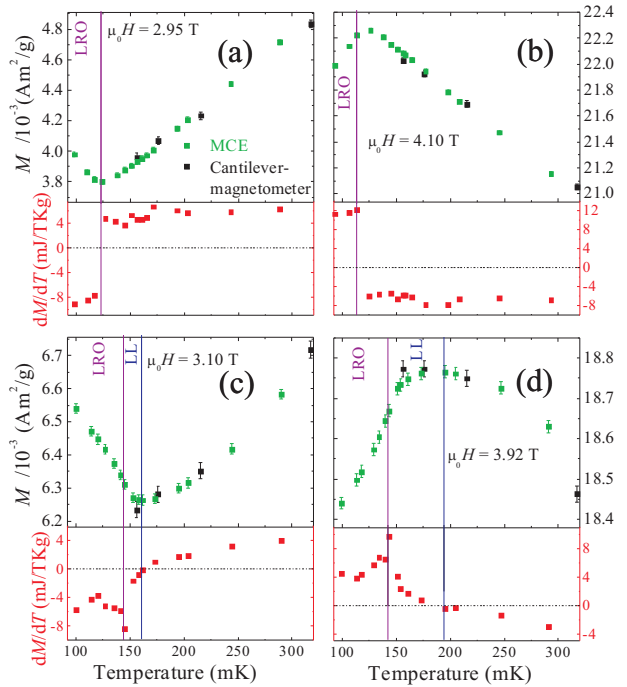}
	\caption[Magnetization of copper nitrate obtained from magnetocaloric effect data]{Magnetization $M(T)$ and the temperature derivative at 2.95\,T (a), 4.1\,T (b), 3.1\,T (c), and 3.92\,T (d) obtained from MCE data. The purple lines indicate the transition into the LRO phase, while the blue lines mark the transition into the LL regime.}
	\label{Figure3}
\end{figure}

The temperature dependent magnetization can be calculated from the MCE via ($-\frac{1}{T} \frac{\delta Q_{\text{MCE}}(H)}{\text{d} (\mu_0 H)}=$ d$M$/d$T$) and integrating the data with respect to temperature. These results are scaled to the cantilever measurements to produce data values on an absolute scale (Fig. \ref{Figure3}).

For a magnetic field of 2.95\,T (Fig. \ref{Figure3}(a)) a pronounced minimum defines the phase transition into the LRO phase. This feature closely resembles the cusp-like minimum in the susceptibility of BEC materials such as TlCuCl$_3$ and marks the boundary for BEC of triplons \cite{Oosawa1999,Nikuni2000}. Further, close to $\mu_0 H_{c2}$, at 4.10\,T (Fig. \ref{Figure3}(b)), a maximum in the magnetization indicates this transition, again in agreement with findings on BEC materials such as BaCuSi$_2$O$_6$ \cite{Jaime2004}. Altogether, the LRO phase in copper nitrate can therefore be interpreted as a BEC phase.

For applied fields closer to the top of the LRO dome, two features are observed. At 3.10\,T (Fig. \ref{Figure3}(c)) a round minimum indicates the crossover into the LL, while a kink defines the transition into the LRO phase. The transition into the LRO phase is very pronounced in the temperature derivative d$M$/d$T$, where a peak defines the transition clearly. Further, the minimum of the magnetization is indicated by a zero crossing of the derivative. For 3.92\,T (Fig. \ref{Figure3}(d)) the situation is similar, but the magnetization minimum is replaced by a maximum. This magnetization evolution with a minimum/maximum for a crossover into a LL regime and a kink for a transition into a 3D ordered phase is in perfect agreement with the predictions by Wessel \textit{et al.} \cite{Wessel2001} for a spin ladder. Altogether, the interpretation of the features in the MCE as phase transitions and crossovers is fully consistent with the observations made for the magnetization $M(T)$.

\begin{figure}[h]
	\centering
		\includegraphics[width=1\linewidth]{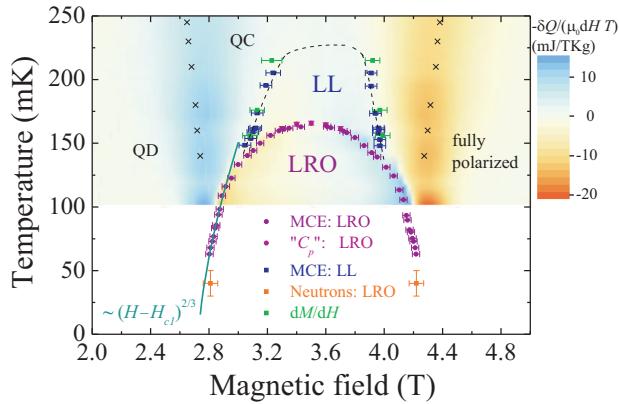}
	\caption[Phase diagram of copper nitrate for magnetic fields along the $b$ direction]{Phase diagram of copper nitrate for applied magnetic fields along the $b$ direction. The dotted line is a guide to the eye. The contour plot represents the data $-\frac{1}{T} \frac{\delta Q_{\text{MCE}}(H)}{\text{d} (\mu_0 H)}$. The crosses indicate local maxima in the heat capacity obtained from simulations using exact diagonalization. They represent crossovers between the QD, QC, and the fully polarized regime; for details see text.}
	\label{Figure4}
\end{figure}

The jumps and zero-crossings from the MCE data and the maxima of the magnetization $M (H)$ data are summarized in a magnetic phase diagram (Fig. \ref{Figure4}). The upper phase boundary of the LRO phase was further defined by jumps in the heat capacity (not shown). As well, the transition fields were observed by neutron diffraction at 40\,mK (not shown). The antiferromagnetic contribution to the nuclear Bragg peak (40$\bar{2}$) indicate the onset of the LRO phase and was added to the phase diagram \cite{neutrons}. Following the presentation in the Refs. \cite{Ruegg2008,Thielemann2009} we include a contour plot of the data $-\frac{1}{T} \frac{\delta Q_{\text{MCE}}(H)}{\text{d} (\mu_0 H)}$.
 
Overall, the general appearance of the phase diagram is similar to that of the spin ladder series (Hpip)$_2$Cu(Br,Cl)$_4$. The phase boundary of the BEC phase forms a dome, which extends up to 166\,mK at 3.52\,T. For a 3D BEC-phase a critical behavior of the phase boundary with the universal critical exponent $\phi=2/3$ is expected \cite{Giamarchi1999}. Accordingly, the phase boundary was fitted to $T_c  \propto(H-H_{c1})^{\phi}$. A fit with $\phi=2/3$ ($H_{c1}=$ 2.73\,T) describes the evolution of the phase boundary within the error bars (blue line in Fig. \ref{Figure4}), fully consistent with the properties of a BEC phase. We note, however, that the phase boundary is rather steep causing an uncertainty in the determination of $\phi$ of about $\pm$0.2. Further, at higher temperatures a 1D LL regime is identified. The LL regime has a dome shape located above that of the LRO phase, with an experimentally determined maximum temperature between 215\,mK and 244\,mK. In contrast to the LRO-phase, the LL dome appears rather asymmetric in shape.

To quantify the experimental data in terms of the model of an alternating antiferromagnetic chain (Eq. (1)), the magnetization, MCE, and heat capacity were calculated for a ring of 14 spins using exact diagonalization (full diagonalization) and for temperatures above the LRO phase using the software ALPS release 2.0 \cite{Bauer2011,Albuquerque2007}. Very good agreement between experimental data and calculation was obtained for temperatures down to 195\,mK using the exchange constants $J_1/k_B=$ 5.10(2)\,K and $J_2/k_B=$ 1.20(2)\,K (calculated data as blue lines in Figs. \ref{Figure1}(a) and \ref{Figure2}). These values agree nicely with those from Xu \textit{et al.} \cite{Xu2000}. We can now further define the phase diagram, in particular the upper boundary of the LL dome. Due to the difficulty in extracting the zero crossing points from the experimental data in this region, we can use the calculated MCE curves to define the upper boundary. It was found at $T_{c,\text{max, LL}}=220(5)$\,mK, which is in full agreement with the experimental data and marked in Fig. \ref{Figure4} as a dashed line. The asymmetry of the phase boundary is derived by the slopes of the MCE data with a steeper phase boundary/slope observed on the high field side. This translates directly into an asymmetry of the entropy peak intensities in which the high field peak maintains a higher entropy compared with the low field peak. In a similar vein the experimental and calculated data d$M$/d$H$ show the same intensity variation as the entropy, with higher intensity at the high field region. Accordingly, this asymmetry was also observed in the ED calculation of our 1D chain model. Altogether it implies that the asymmetry is intrinsic to this system and may in fact be a fundamental property in alternating chain LL regimes. Next, following Ref. \cite{Ruegg2008}, from the local maxima of the calculated heat capacity we distinguish between the {\it quantum disordered} (QD), {\it quantum critical} (QC) and {\it fully polarized} regimes of the phase diagram (Fig. \ref{Figure4}). The QD phase is characterized by a well defined gap between singlet and triplet states, while the QC phase shows a collapse of this gap. In this part of the phase diagram the triplet state becomes populated with no long-range magnetic correlations. Finally, in the fully polarized regime a gap reopens and the moments are constrained to align along the field.

For the spin ladder series (Hpip)$_2$Cu(Br,Cl)$_4$ the low temperature properties were accounted for by the rung and leg couplings, $J_{rung}$ and $J_{leg}$, plus the residual interchain coupling $J'$ \cite{Klanjsek2008,ward2013,bouillot2011,Thielemann2009,Ruegg2008}. It was argued that for the series (Hpip)$_2$CuBr$_4$ to (Hpip)$_2$CuCl$_4$ the energy scales of the LL and the BEC phase scale with the ladder coupling $J_{leg}$ and $J'$, respectively. With the coupling strengths $J_{rung}$;$J_{leg}$ for (Hpip)$_2$CuBr$_4$ (12.6\,K; 3.55\,K) and (Hpip)$_2$CuCl$_4$  (3.42\,K; 1.34\,K) the coupling ratio $\gamma = J_{leg} / J_{rung}$ is similar (0.28 (Br$_4$) and 0.39 (Cl$_4$)). (Hpip)$_2$CuBr$_4$ exhibits field induced long range magnetic order below $\sim$100\,mK, corresponding to an interchain coupling $J'$ of the order of a few 10s of mK. For (Hpip)$_2$CuBr$_4$ ((Hpip)$_2$CuCl$_4$) a LL is observed up to 1.4 K (0.5 K), implying that the maximum of the LL crossover dome lies at $\sim 0.4 J_{leg}$. 

In comparison, for the alternating chain copper nitrate we find for the quantity analogous to $\gamma$, the ratio $J_2 / J_1 = 0.24$, {\it i.e.}, a similar degree of dimerization. As well, the residual 3D coupling strength is of the order of a few 10s of mK. In contrast, Luttinger liquid behavior is observed over a much smaller temperature range, that is up to 220\,mK $\approx  0.18 J_2$. Thus, while copper nitrate structurally and in terms of magnetic coupling strengths is a material even closer to the 1D limit than (Hpip)$_2$Cu(Br,Cl)$_4$, 1D physics is more pronounced in the latter compound. Here, it is intriguing to compare the ladder Hamiltonian
\begin{multline}
H= \sum_{i} \left(J_{rung} \boldsymbol{S}_{i,1} \boldsymbol{S}_{i,2} + J_{leg} \left( \boldsymbol{S}_{i,1} \boldsymbol{S}_{i+1,1} + \boldsymbol{S}_{i,2} \boldsymbol{S}_{i+1,2} \right) \right)\\
 -h \sum_{i} S_i^z
\end{multline}
with the alternating chain Hamiltonian Eq.(1). It is tempting to associate $J_1$ with $J_{rung}$ and $J_2$ with $J_{leg}$. Within this picture, we speculate that on a qualitative level the more pronounced 1D behavior of the spin ladder might result from the additional 1D exchange path provided by the ladder structure.

In conclusion, this work shows for the first time a detailed mapping of both LL and LRO phases in an alternating antiferromagnetic chain. While the alternating chain, copper nitrate, shows similar features to the ladder compounds, a clear difference in the energy scales is revealed. Notably, we have been able to use the single technique of MCE to define both the 1D LL and 3D LRO phase boundaries. This represents the clearest evidence of a dimensional crossover available in the literature to date and may motivate further investigations of this fascinating observation. 

We gratefully acknowledge fruitful discussions with W. Brenig and A. Honecker. Further, we acknowledge access to the experimental facilities of the Laboratory for Magnetic Measurements (LaMMB) at HZB. This work has partially been supported by the DFG under Contract Nos. WO 1532/3-1 and SU229/9-1.

\end{document}